\documentclass{ws-ijmpd}

\usepackage{times}
\usepackage{graphicx}
\usepackage{subfigure}

\newcommand{\beq}{\begin{equation}}
\newcommand{\eeq}{\end{equation}}

\newcommand{\ba}{\begin{eqnarray}}
\newcommand{\ea}{\end{eqnarray}}

\newcommand{\bfd}{\mathbf{d}}

\newcommand{\bfP}{\mathbf{P}}

\def\gs{\mathrel{\lower0.6ex\hbox{$\buildrel {\textstyle >}\over{\scriptstyle \sim}$}}}
\def\ls{\mathrel{\lower0.6ex\hbox{$\buildrel {\textstyle <}\over{\scriptstyle \sim}$}}}

\newcommand{\aap}{A\&A}
\newcommand{\apj}{ApJ}
\newcommand{\apjl}{Astrphys. J. Lett.}

\newcommand{\aj}{AJ}
\newcommand{\nat}{Nature}
\newcommand{\prd}{Phys. Rev. D}
\newcommand{\mnras}{MNRAS}
\newcommand{\physrep}{Phisycs Rep.}

\begin{document}

\markboth{Radicella, Sereno, Tartaglia}
{Constraining CD theory}

%
\catchline{}{}{}{}{}
%

\title{Cosmological constraints for the Cosmic Defect theory}

\author{Ninfa Radicella}

\address{Departament de F\'{\i}sica, Universitat Aut\`{o}noma de Barcelona\\
ninfa.radicella@uab.cat}

\author{Mauro Sereno}

\address{Dipartimento di Fisica, Politecnico di Torino, Corso Duca degli Abruzzi 24\\
Torino, 10129, Italia\\
and\\
INFN, Sezione di Torino, Via Pietro Giuria 1\\
Torino, 10125, Italia\\
mauro.sereno@polito.it}

\author{Angelo Tartaglia}

\address{Dipartimento di Fisica, Politecnico di Torino, Corso Duca degli Abruzzi 24\\
Torino, 10129, Italia\\
and\\
INFN, Sezione di Torino, Via Pietro Giuria 1\\
Torino, 10125, Italia\\
angelo.tartaglia@polito.it}

\maketitle

\begin{history}
\received{Day Month Year}
\revised{Day Month Year}
\comby{Managing Editor}
\end{history}

\begin{abstract}
The Cosmic Defect theory has been confronted with four observational constraints: primordial nuclear species abundances emerging from the big bang nucleosynthesis; large scale structure formation in the universe; cosmic microwave background acoustic scale; luminosity distances of type Ia supernovae.
The test has been based on a statistical analysis of the a posteriori probabilities for three parameters of the theory. The result has been quite satisfactory and such that the performance of the theory is not distinguishable from the one of the $\Lambda$CDM theory. The use of the optimal values of the parameters for the calculation of the Hubble constant and the age of the universe confirms the compatibility of the Cosmic Defect approach with observations.
\end{abstract}

\keywords{Space-time; Cosmology; Extended theories of gravity.}

\section{Introduction}	

From the very beginning of the theory of relativity both in its special form (SR) and in its general form (GR) a problem of interpretation of the nature of space-time has been present. This problem, in most cases, is implicit rather than explicit, but it is there. Is space-time a sort of field?  which would be perfect for people trying to quantize the gravitational interaction. This cannot be, however, since, in general, fields are described \emph{in} space-time; what would then be the background of space-time? What in any case is granted is that space-time is indeed something physical, not a mere mathematical artifact. Space-time interacts with matter as the Einstein equations tell us, so even not being matter it has properties on its own.

On another side we know that space-time is a curved four-dimensional manifold endowed with Lorentzian signature, and we also know that space, at the cosmic scale, appears to be expanding with a typical symmetry: the Robertson-Walker (RW) symmetry. Even more: some pieces of observation tell us that the expansion is accelerated \cite{perl,riess}. This acceleration is not evidently due to matter, so commonly it is ascribed to ``something else" which is dubbed in various ways according to different theories, but is mostly known as \emph{dark energy} \cite{den,ein,eina,zla,car,cal,vik,zhang,li,ark,mukh}. What is dark energy? There are as many answers as there are theories, but it is difficult to say that the situation is clear and that each interpretation of dark energy is physically well motivated.

We refer here to the Cosmic Defect (CD) theory. It describes space-time on the basis of the existing analogy with usual three-dimensional elastically deformable continua; the classical theory is of course extended to four dimensions and endowed with the appropriate signature. All features of GR are preserved. The theory is exposed in \cite{ta+ra10} and will be shortly reviewed in the next section. There are two relevant features with CD: a) the global symmetry of the universe is due to a defect in the texture of space-time working as defects do in material continua; b) the deformation of the manifold, due both to defects and to the presence of matter is expressed by a strain tensor which coincides with the non-trivial part of the metric tensor. To the strain tensor an \emph{elastic} potential energy density corresponds, whose presence affects the global behavior of space-time and is responsible for the accelerated expansion. The theory is in our view appealing, but of course it must be tested on facts. We already used it to work out the luminosity/distance curve of type Ia supernovae with good results \cite{ta+ra10}, however the fit of the luminosity of SnIa's is not difficult to be obtained by a variety of different theories so it is not really discriminating among them; this is the reason why we decided to test CD against more constraints. One such constraint concerns the cosmological nucleosynthesis (Big Bang Nucleosynthesis: BBN): the expansion law of the universe must be such that the appropriate pressure and temperature conditions last long enough (and not more) to produce He and D in the percentages we observe or deduce from indirect observations \footnote{More light elements would enter a more refined and detailed analysis but the two considered in the text are enough for our purposes.}.

We considered two other main tests: the large scale structure (LSS) formation, which probes the universe history at the time of matter-radiation equality, and the acoustic scale of the cosmic microwave background (CMB), which gives an independent constraint at the recombination era. This allowed us to compare the performance of the CD theory with other theories and especially with $\Lambda$CDM.
After exposing the method that we have adopted in order to validate the constraints, we shall verify also the acceptability of the estimates obtained from CD for the Hubble constant and the age of the universe. We shall see once more that the results are more than acceptable and we shall discuss them in the concluding section adding some consideration regarding the Solar system environment.

\section{Review of the CD theory}
\label{sec_theory}

The core idea of the Cosmic Defect theory is that the actual space-time manifold with its global and local curvature behaves as a four-dimensional elastic continuum, so that one may think that the natural situation is obtained introducing strain in an initially flat (Minkowski) manifold. The properties of the strained manifold are expressed in terms of two parameters, which are the Lam\'e coefficients of space-time, $\lambda$ and $\mu$. The details of the theory may be found in \cite{ta+ra10}; here we only recollect that the global RW symmetry is assumed to be induced by a cosmic defect and that the actual metric tensor is composed of two contributions:
\beq
\label{metrica}
g_{\mu\nu}=\eta_{\mu\nu}+2\epsilon_{\mu\nu},
\eeq
where $\eta_{\mu\nu}$ is the Minkowski metric tensor and $\epsilon_{\mu\nu}$ is the strain tensor.
According to the analogy with a deformed elastic material we expect the strain to be associated with an elastic deformation energy written as:
\beq
\label{energy}
W_{e}= \frac{1}{2} \lambda \epsilon^2+\mu\epsilon_{\alpha\beta}\epsilon^{\alpha\beta}.
\eeq
Now $\epsilon=\epsilon^{\alpha}_{\alpha}$ is the trace of the strain tensor.
The energy in (\ref{energy}) allows to write a new action integral including both space-time and matter/energy in the form
\beq
\label{action}
S=\int{(R+ \frac{1}{2} \lambda \epsilon^2+\mu\epsilon_{\alpha\beta}\epsilon^{\alpha\beta}+L_\mathrm{matter})\sqrt{-g}d^4x}.
\eeq
$L_\mathrm{matter}$ is the Lagrangian density of ordinary matter/energy and the rest has the usual meaning. The ``elastic" potential energy term belongs to geometry, i.e. space-time, even though it looks like some matter contribution. Considered from the field theoretical viewpoint this new contribution implies a "mass" associated with the gravitational interaction; usually this is said as the graviton being massive, which fact has relevant consequences when studying the propagation of gravity and, in particular, of gravitational waves \cite{viss,dam,def,ark}. In any case our conceptual framework is entirely classical so that, properly speaking, there are no gravitons, but rather it turns out that gravity has a finite range \cite{boul,baba}. We are not discussing the issue in this paper; we only remark that the numerical values we find after our tests and fits to observation tell us that the effects of the space-time strain are totally negligible at the scale of the Solar system.

As we have seen, everything depends on the strain tensor, which in turn depends on the way events on the reference manifold are associated to their corresponding events in the natural manifold. Actually there are infinite possible ways to get a given final situation starting from a flat initial one. This apparent freedom of choice has indeed a physical meaning since our manifolds are physical. So different choices correspond to different strains and the Hamilton principle permits to identify the least strain configuration. In practice what can be seen as a gauge freedom appears as a gauge function in the line element of the flat reference written using the coordinates of the curved natural manifold. We expect the different gauge choices not to affect the general laws, such as the expansion law in the cosmological case, rather to point out different properties of space-time, i.e. different values of its Lam\'e coefficients.

Applying all the above to the RW symmetry in (\ref{action}) one is left with two free functions. One of them is physical: the scale factor $a(\tau)$, depending on the cosmic time $\tau$. The other is the mentioned gauge freedom expressing the difference between cosmic time and the time coordinate on the flat reference manifold; in other words, using the cosmic coordinates, the flat line element of the reference manifold is written

\beq
\label{flat}
ds^{2}= b^2(\tau)d\tau^2-dx^2-dy^2-dz^2
\eeq

and $b(\tau)$ is the mentioned gauge function.

From the RW line element and from (\ref{flat}) it is straightforward to write the non-zero elements of the strain tensor, which are
\ba
\epsilon_{00}= \frac{1-b^2}{2} \nonumber \\
\epsilon_{xx}= \epsilon_{yy}= \epsilon_{zz}= \frac{1-a^2}{2} \nonumber  \\
\ea

Introducing the explicit form of the strain tensor into the Lagrangian of our problem, then applying the usual variational principle, one arrives to a pair of equations: one is solved to fix the gauge i.e. to give the explicit dependence of $b$ on $a$; the other, after one integration, gives the expansion law of the universe, which is implicitly written in the formula for the Hubble parameter \cite{ta+ra10}:
\ba
\label{hubble0}
H & = &\frac{\dot{a}}{a}= c\sqrt{\frac{B}{16}} \left\{ 3\left(1-\frac{(1+z)^2}{a_0^2}\right)^2 \right. \nonumber \\
& +& \left. \frac{8\kappa }{3B}(1+z)^3\left[\rho_\mathrm{m0}+\rho_\mathrm{r0}(1+z)\right] \right\}^{1/2} .
\ea
Dots are derivatives with respect to cosmic time; $z$ is the redshift factor; $a_{0}$ is the value of the present cosmic scale factor; $\kappa = 16\pi G/c^2$ is the coupling constant between matter/energy and geometry; $\rho_\mathrm{m0}$ is the present mass density of the universe and $\rho_\mathrm{r0}$ is the present radiation energy density. The $B$ parameter is the bulk modulus of space-time and its dependence on the Lam\'e coefficients is obtained from the gauge condition. It is:
\beq
\label{bulk}
B=\frac{\mu}{4}\frac{2\lambda+\mu}{\lambda+2\mu}
\eeq
that works for any combination of the Lam\'e coefficients excepting $\lambda=-2\mu$ \footnote{The expression found in \cite{ta+ra10} has been obtained forcing the gauge to be $b=1$ which is only consistent with the condition $\lambda=0$.}.

In the approach corresponding to (\ref{hubble0}) the content of the universe is simply dust and radiation.
Formula (\ref{hubble0}) is the starting point for the elaborations in the rest of the paper.

\section{Observational constraints}
\label{sec_obse}

Observational cosmology allows us to put strong constraints on the expansion history of the universe over a very large span of time. Here, we want to test if the scale factor evolution in (\ref{hubble0}) is compatible with such history. In order to perform a conservative analysis, we looked at geometrical tests based on a minimum set of hypotheses. We did not consider evidence more heavily dependent on the fraction of baryonic matter (such as peaks in the galaxy power spectrum) or on the present linear-theory mass dispersion $\sigma_8$. Constraints from the cosmic microwave background (CMB), which require modeling of additional parameters for spatial curvature, amplitude and slope for the tensor spectrum and optical depth to last scattering, were not considered too. We have however estimated the $CMB$ acoustic scale which can show whether our theory behaves differently from $\Lambda$CDM and other theories or not.

As standard relativistic contributions, we considered a photon background at $T=2.728~\mathrm{K}$ \cite{fix+al96} together with three species of massless neutrinos.

\subsection{Nucleosynthesis}

At very early time, the matter/energy budget is radiation dominated and the Hubble factor can be approximated as
\beq
\label{nucl1}
H^2 \simeq c^2 \frac{\kappa}{6}\left( 1+\frac{B}{B_{a_0}}\right)\rho_\mathrm{r0}z^4,
\eeq
with
\beq
\label{nucl2}
B_{a_0}\equiv \frac{8}{9}\kappa \rho_\mathrm{r0}a_0^4.
\eeq
The strain energy is not supposed to affect the cross-sections of local nuclear interactions, whereas, as far as the expansion rate is considered, it effectively boosts the radiation density by a factor $X_\mathrm{Boost}=(1+B/B_{a_0})$. Then, the strain of space-time enters nucleosynthesis processes only through a modified expansion. Due to the increase in the overall rate of expansion, neutron freeze-out happens earlier, raising the final helium abundance. Non-standard boost factors ($X_\mathrm{Boost}\neq 1$) can then be tightly constrained by fitting measured abundances.

Several methods have been proposed to determine the $^4He$ mass fraction, $Y_p$ \cite{ioc+al09} with systematic errors being the main responsible for the spread determination. To account for this, \cite{ioc+al09} performed an analysis of several values in literature, ending up with an estimate of $Y_p=0.250\pm0.003$. This range of values is still fully compatible with other more recent estimates \cite{iz+th10}, so we still refer to the conservative analysis in \cite{ioc+al09}. The related constraint on the boost factor is $X_\mathrm{Boost}= 1.025\pm 0.015$ \cite{ioc+al09}, from which we can write a $\chi^2$ constraint as
\beq
\label{nucl3}
\chi^2_\mathrm{BBN} = \left( \frac{X_\mathrm{Boost}-1.025}{0.015}\right)^2 .
\eeq

\subsection{Structure formation}

Matter perturbations can not grow in a universe expanding as a radiation-dominated background. The space-time strain effectively increases the radiation-like density and matter dominance is postponed. The effective boost $X_\mathrm{Boost}$ affects the scale of the particle horizon at the equality epoch, $z_\mathrm{eq}\simeq 3150$ \cite{kom+al10}.  On the other hand, the Newtonian limit of gravity is still fulfilled in presence of defects \cite{ta+ra10}, so that, in a CD model, perturbation growth is affected mainly through the modified expansion rate of the background. The horizon at the equality is imprinted in the matter transfer function. The constraint from large scale structure (LSS) becomes \cite{pea99}:
\beq
\label{lss1}
\left( \Omega_\mathrm{m0}h\right)_\mathrm{apparent} = X_\mathrm{Boost}^{-1/2} \left( \Omega_\mathrm{m0}h\right)_\mathrm{true} .
\eeq
where $h$ is the Hubble constant $H_0$ in units of $100~\mathrm{km~s}^{-1}\mathrm{Mpc}^{-1}$ and $\Omega_\mathrm{m0}$ is the matter density in units of the critical density $\rho_\mathrm{Cr}\equiv 3 H_0^2/(8\pi G)$.

We considered the final analysis from the 2dF Galaxy Redshift Survey \cite{col+al05} which found $\left( \Omega_\mathrm{m0}h\right)_\mathrm{apparent}= 0.168 \pm 0.016$. The fitting procedure in \cite{col+al05} was performed trying to limit the number of parameters so that you can reliably use their result combined with Eq.~(\ref{lss1}) to get constraints on extra-physics. We share the same assumption on the index of the primordial power spectrum ($n=1$). The related constraint on the cosmological parameters of the CD theory can then be written as
\beq
\label{lss2}
\chi^2_\mathrm{LSS} = \left( \frac{ X_\mathrm{Boost}^{-1/2}\Omega_\mathrm{m0}h -0.168}{0.016} \right)^2 .
\eeq

\subsection{Cosmic Microwave Background}

The temperature power spectrum of CMB is sensitive to the physics at the decoupling epoch, $z_\mathrm{LS} \sim 1090$, as well as the expansion history between now and the last scattering surface. Some of the main features of the spectrum can be summarized by a number of parameters. Among them, we consider the acoustic scale $l_\mathrm{A}$\cite{hu+su96,kom+al10},
\beq
l_\mathrm{A}=(1+z_\mathrm{LS})\pi \frac{D_\mathrm{A}(z_\mathrm{LS})}{r_\mathrm{s}(z_\mathrm{LS})},
\eeq
where $r_\mathrm{s}$ is the sound horizon at recombination and $D_\mathrm{A}$ is the angular diameter distance to the last scattering surface. The cosmological model we are testing affects the power spectrum in two main ways. Firstly, the radiation boost $X_\mathrm{Boost}$ affects the location of the acoustic peaks, which depends on the expansion factor at the matter-radiation equality, $a_{eq}$. The formula for the sound horizon is then the same as for the $\Lambda$CDM model \cite{hu+su96,el+mu07} with the only caveat that now $a_{eq}=X_\mathrm{Boost} \rho_{r0}/\rho_{m0}$. Secondly, the angular diameter distance depends on the expansion history,
\beq
D_\mathrm{A}(z_\mathrm{LS})=\frac{c}{(1+z_\mathrm{LS})}\int_0^{z_\mathrm{LS}}\frac{d z}{H(z)}.
\eeq
The values determined for $l_A$ are quite model-independent, so that we can consider the latest results from WMAP-7 \cite{kom+al10}. We take $l_A^{Obs} =302.69 \pm 0.76 \pm 1.00$, where the first error is the statistical error in a $\Lambda$CDM model and the second error gives an estimate of the uncertainty connected to the model \cite{el+mu07}. Since these errors are independent they can be added in quadrature. The statistical constraint then reads
\beq
\chi^2_{CMB} = \left( \frac{l_A (B,\ \Omega_\mathrm{m0},\ B_{a_0})-302.69}{1.26} \right)^2 .
\eeq

\subsection{Supernovae}

Supernovae (SNe) of type Ia are supposed to be the best tracers of the recent expansion history of the universe. \cite{kow+al08} provided a sample of 307 SNe  for cosmological analysis. As usual, we can compare the measured to the predicted distance moduli,
\beq
\label{sne1}
d(z)=25+5\log_{10} \left[(1+z) \int_0^z \frac{(c/\mathrm{Mpc})}{H(z')}dz'\right].
\eeq
The related $\chi^2$ term is
\beq
\label{sne2}
\chi^2_\mathrm{SNe}=   \sum_i \left( \frac{d_i-d(z_i )}{\Delta d_i} \right)^2 ,
\eeq
where $d_i$ is the measured distance modulus at $z_i$.

\section{Statistical analysis}

\begin{figure*}
	\centering
		\subfigure{\includegraphics[width=0.3\textwidth]{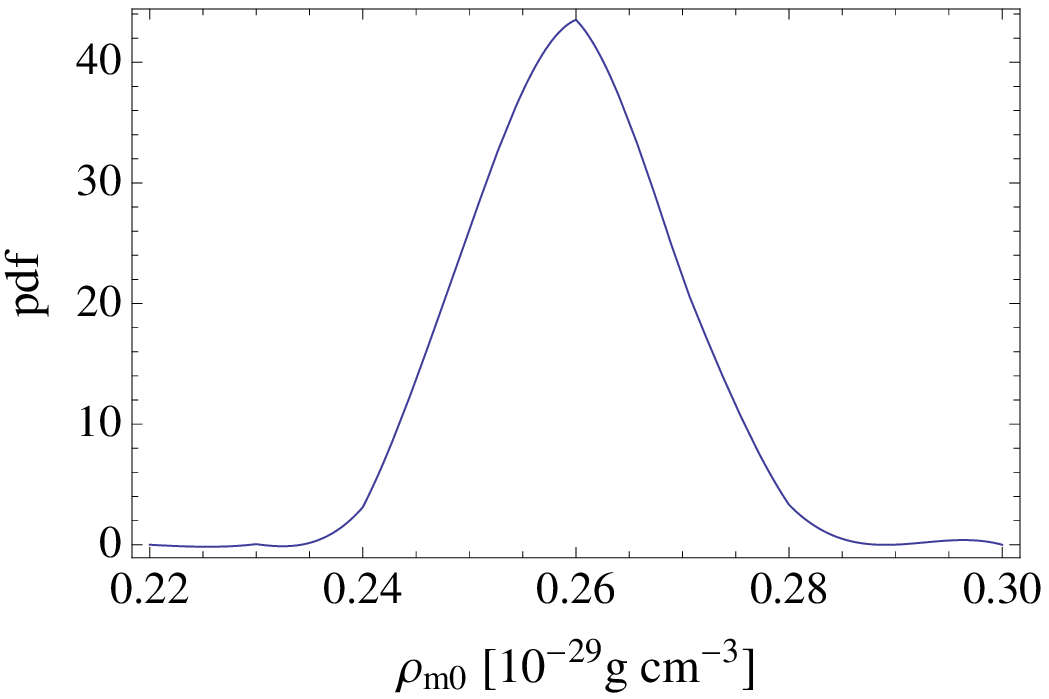}}
		\subfigure{\includegraphics[width=0.3\textwidth]{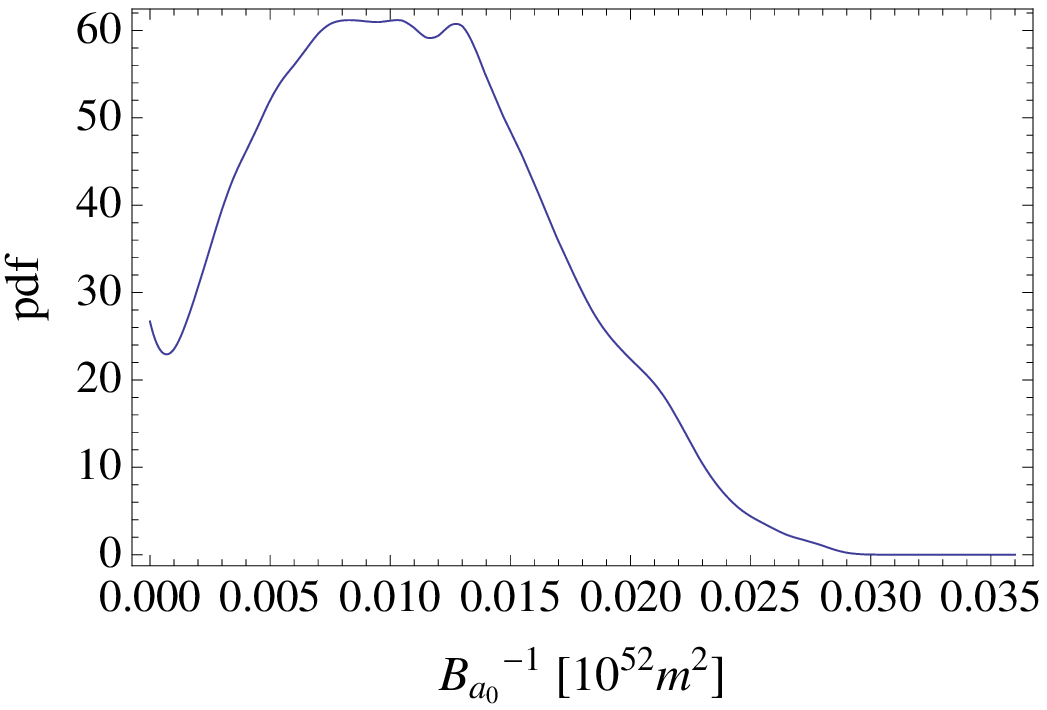}}
		\subfigure{\includegraphics[width=0.3\textwidth]{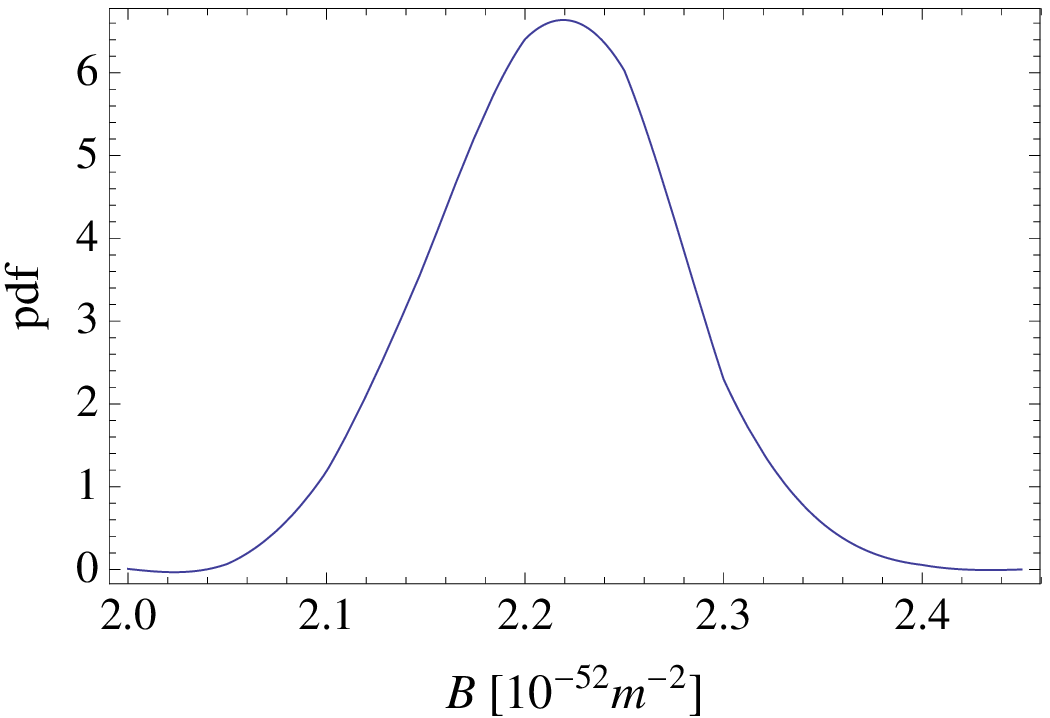}}
	\caption{Posterior probability density functions for the parameters $\rho_\mathrm{m0}$ (left panel), $B_{a_0}^{-1}$ (middle) and $B$ (right). Units are as in Table~\ref{tab_par}.}
	\label{fig_pdf}
\end{figure*}

\begin{table}
\tbl{Estimated parameter values. The maximum likelihood estimates are reported in brackets.}
{
\begin{tabular}{c|c|c}
        \hline
        \noalign{\smallskip}
	$B~(10^{-52}\mathrm{m}^{-2})$	&$\rho_\mathrm{m0} (10^{-29}\mathrm{g\ cm}^{-3})$	 &$B_{a_0}^{-1} (10^{52}\mathrm{m}^2)$  \\
	\hline
        \noalign{\smallskip}
	$2.22~(2.22)\pm 0.06$ &	$0.260~(0.258) \pm 0.009$ &	$0.011~(0.009) \pm 0.006$ \\
        \noalign{\smallskip}
\hline
\end{tabular}
\label{tab_par}
}
\end{table}

We performed the statistical analysis with standard Bayesian methods \cite{le+br02,mac03}. The Bayes theorem states that
\beq
\label{baye}
p(\bfP | \bfd) \propto {\cal L}( \bfP|\bfd) p(\bfP),
\eeq
where $p(\bfP | \bfd)$ is the posterior probability of the parameters $\bfP$ given the data $\bfd$, ${\cal L}( \bfP|\bfd)$ is the likelihood of the data given the model parameters and $p(\bfP)$ is the prior probability distribution for the model parameters.

For our analysis, three model parameters are enough to describe the CD theory. The discussed constraints do not require us to distinguish between baryonic and dark matter, so that in the following analysis we can consider a single parameter density for the dust-like matter, $\rho_\mathrm{m0}$. Cosmic defects characterize the $B$ parameter. Finally, the present value of the scale factor is described in terms of $B_{a_0}$. For convenience, we actually used as parameter the inverse of $B_{a_0}$. Finally, $\bfP=\left\{ \rho_\mathrm{m0},\ B, \ B_{a_0}^{-1}\right\}$.

In order to investigate the parameter space for the CD model, we have to write the likelihood. The constraints discussed in Section~\ref{sec_obse} allow us to write the total $\chi^2$ as
\beq
\chi^2(\rho_\mathrm{m0},B, B_{a_0}^{-1}|\bfd)  =  \chi^2_\mathrm{SNe} + \chi^2_\mathrm{BBN} +  \chi^2_\mathrm{LSS}  + \chi^2_\mathrm{CMB} \label{stat1}.
\eeq
The related likelihood is then ${\cal L} \propto e^{-\chi^2/2}$.

As a priori information, we considered flat distributions for each parameter. The relativistic energy density was fixed at $\rho_\mathrm{r0} \simeq 7.8\times10^{-34}~\mathrm{g/cm}^3$.

The parameter space was explored with Monte Carlo Markov chains methods. We run four chains, each one with $10^4$ samples. Convergence criteria are safely satisfied, with the Gelman and Rubin ratio \cite{ge+ru92,le+br02} being $\ls 1.003$ for each parameter. Results are summarized in Fig.~\ref{fig_pdf} and Table~\ref{tab_par}.

The posterior probability densities (pdfs) are plotted in Fig.~\ref{fig_pdf}. Distributions have been smoothed with a Gaussian adaptive kernel with reflective boundary conditions \cite{vio+al94,ryd96}. The pdfs are very regular and single peaked. Parameter estimates, computed as mean and standard deviation of the final marginalized distributions, are reported in Table~\ref{tab_par}. We also performed a standard maximum likelihood analysis. The maximum was found with a downhill simplex algorithm \cite{pre+al92}. As expected for single-peaked distributions, Bayesian estimates coincide almost perfectly with results from the maximum likelihood analysis.

The elastic parameter $B$ is mainly determined by late time evolution, when the space-time strain energy propels acceleration, whereas $B_{a_0}$ is pretty much constrained by the BBN, the constraint from LSS being weaker. The matter density is well determined thanks to LSS and SNe. The density parameter reads $\Omega_\mathrm{m0}=0.28\pm 0.01$.

Today's scale parameter is not well constrained by the data. LSS and BBN require $a_0$ to be large enough so that the boost factor $X_\mathrm{boost}$ is small at early times. We got $a_0>19.9$ at the 99\% confidence level (CL) and $X_\mathrm{boost} = 1.024 \pm0.013$ with $X_\mathrm{boost}<1.053$ at the 99\% CL. On the other hand, SNe data are compatible also with much smaller values of order unity for $a_0$. Note that the tail at large values for the a posteriori pdf of $a_0$ is strongly dependent on the chosen prior on $B_{a_0}^{-1}$, whereas the lower limit is not.

\section{Compatibility checks}
\label{sec:compat}

\subsection{Some observational predictions}
\label{subsec:one}

To further test the performance of the cosmic defect theory, we can test some predictions based on the above statistical analysis. The inferred Hubble parameter turns out to be $H_0=70.2 \pm 0.5~\mathrm{km~s}^{-1}\mathrm{Mpc}^{-1}$, in very good agreement with the local estimate of $73 \pm 2\pm4~\mathrm{km~s}^{-1}\mathrm{Mpc}^{-1}$ based on high precision distance determination methods,  \cite{fr+ma10}. The estimated age of the universe is $t_0=13.7\pm0.1~\mathrm{Gy}$ in good agreement with lower limits obtained by the ages of the oldest globular clusters ($12.6_{-2.2}^{+3.4}~\mathrm{Gy}$ at 95\% CL \cite{kr+ch03}) and radioactive dating ($12.5\pm 3~\mathrm{Gy}$ \cite{cay+al01}).


\subsection{Comparison to  $\Lambda$CDM}
\label{subsec:two}

\begin{table}
\tbl{$\chi^2$ and $BIC$ values for the CD model versus $\Lambda$CDM.}
{
\begin{tabular}{ccccc}
        \hline
        \noalign{\smallskip}
				&	 \multicolumn{2}{c}{$\chi^2$}	&	\multicolumn{2}{c}{$BIC$} \\
	Constraints	&	CD		&	$\Lambda$CDM 	&	CD		&	$\Lambda$CDM \\
	\hline
        \noalign{\smallskip}
	SNe 					&	$310.3$	&	$311.9$		&	$327.5$		& 	$323.4$	\\
	SNe+LSS				&	$311.9$	&	$313.8$ 		&	$329.1$		&	$325.3$	\\
	SNe+LSS+CMB		&	$314.1$	&	$314.1$		&	$331.3$		&	$325.6$		\\
	SNe+LSS+BBN			&	$312.3$	&	--			&	$330.8$		&	--		\\
	SNe+LSS+BBN+CMB	&	$315.0$	&	--			&	$332.3$		&	--		\\
        \noalign{\smallskip}
\hline
\end{tabular}
\label{tab_chi}
}
\end{table}

CD model predictions for $\Omega_\mathrm{m0}$ and $H_0$ are in good agreement with results for a flat $\Lambda$CDM model \cite{la+li10}.
Let us now discuss how these two competing models compare under a statistical point of view. The quickest way to model comparison is to exploit some information criterium which takes into account how well a model fits the data versus its complexity \cite{lid07}. In particular, the Bayesian information criterium is defined as $BIC = \chi^2 + N_{par}\log N_{const}$, where $\chi^2$ is the total $\chi^2$ for the model, $N_{par}$ is the number of parameters and $N_{const}$ is the number of observational constraints. The best model minimizes the BIC.

The CD model has three free parameters; two parameters, $\{ H_0,~\Omega_\mathrm{m0}\}$, characterize the flat $\Lambda$CDM model. $\chi^2$ performances are
listed in Table~\ref{tab_chi}. Comparison was performed under hypotheses favorable to $\Lambda$CDM. We did not use the BBN constraint, since the studied $\Lambda$CDM model lacks of any additional relativistic species to fit the abundances. This circumstance further penalizes the model with the larger number of parameters. However, $\chi^2$ and $BIC$ values are too close to prefer one model over another.

A full treatment would require the computation of Bayesian evidences. However, due to the very similar likelihood functions, the estimated evidences would be highly dependent on the assumed priors, which is not really informative.

\section{Conclusions}
\label{sec:conc}
We have submitted the CD theory to a consistency test with respect to three main cosmological constraints: primordial nuclear isotopic abundances; large scale structure formation in the universe; luminosity distances of type Ia supernovae. One relevant quantity, the sound horizon, deducible from the CMB anisotropy data has also been evaluated. According to the CD theory space-time is endowed with a strain energy density, whose presence in the Lagrangian density of space-time affects the expansion law of the universe and produces the present accelerated expansion. In early times the effect of the strain term looks like the one of radiation, however it shows up only in the expansion rate without directly affecting either the interaction between radiation (photons and neutrinos) and particles or the cross sections of the nuclear reactions. In practice the relevance of the parameters of the CD theory appears in a boost factor. Similar considerations hold for the large scale structures formation as well as for the sound horizon of the CMB at last scattering. Later, the accelerated expansion, which shows up as a dimming of the SnIa's, is essentially determined by the bulk modulus of space-time treated as an elastic four-dimensional manifold with Lorentzian signature. The constraints posed by observation may be translated into constraints for the values of three parameters of the CD theory. On these bases a statistical analysis has been done on the a posteriori probabilities of the values of the parameters. By these means we have found: a) that CD is compatible with the observational constraints; b) that the results are as good as the ones obtained from the $\Lambda$CDM theory. These conclusions are strengthened by the use of the optimal values of the parameters for determining the Hubble constant and the age of the universe: the values we obtain are in good agreement with the currently accepted ones.

The tests we performed on CMB and LSS implicitly assumed a Newtonian approach for the density perturbation growth. We can check a posteriori that such assumption holds. The LSS test in Sect.~\ref{sec_obse} was based on the main bend visible in the matter transfer function, which arises because the universe expansion is radiation dominated at early times whereas fluctuations in the matter can only grow if dark matter and radiation fall together. Similar considerations stand for the position of the acoustic peak in the CMB spectrum. Both tests then exploit a distinctive geometric feature in the respective spectra.

As for the expansion law of the universe, we used the exact result of the CD theory. For the gravitational potential, we assumed a locally Newtonian potential, i.e., a solution of the standard Poissonian equation. This is fully justified given the size of the Lam\'e coefficients of space-time, which determine the relevant scale for the CD theory.

The test with the SNe is independent of the growth of perturbations and relies only on the expansion history. It constrains $B$, whose size in turn gives that of $\lambda$ and $\mu$. The SNe analysis unambigously shows that the order of magnitude of the Lam\'e coefficients of space-time must be $\sim 10^{-52}\mathrm{m}^{-2}$, see also \cite{ta+ra10}. On a dimensional analysis, local deviation from the Newtonian potential near a mass $m$ must be of the kind of $\lambda r^2$, $\mu r^2$ or $\lambda m r$, $\mu m r$ or higher order corrections. Even on the present scale of a galaxy cluster ($m\sim 10^{14}M_\odot$, $r \sim 1~\mathrm{Mpc}$), the Newtonian term $m/r$ is a couple of orders of magnitude larger than the CD correction. It is then safe to consider a standard Newtonian potential in the test we performed. The basic version of our tests on CMB and LSS does not then require a full treatment of perturbation growth in the CD theory.

Even if the CD theory provides a novel theoretical framework, the above considerations are the same generally done for the cosmological constant. Despite the very different origin, $\Lambda$ is estimated to be of the order of $\sim 10^{-52}\mathrm{m}^{-2}$ too. Its contribution to the local potential is neglected in first approximation when considering density perturbations and the simplest version of the tests on CMB and LSS\cite{col+al05}.

One more remark we can make concerns the possible Solar System effects of the strain of space-time. A full discussion of this issue requires the explicit solution of the Einstein equations with strain in a spherically symmetric stationary case, which is beyond the scope of the present paper. Let us again compare $B$ times the square of the distance from the center of the system to the equally dimensionless quantity $m/r$ that tells us how strong are the gravitational effects for instance in the Solar system. In the environment of any stellar system and for distances of a few astronomical units $m/r$ is of the order of $\sim$ $10^{-10}\div 10^{-8}$. Considering the result in Table~\ref{tab_par} we see that, in the same distance range, it is $Br^2 \sim 10^{-28}$, i.e. some 20 orders of magnitude below the Newtonian mass terms. For this reason we expect the effects of the strain of space-time to show up at cosmic scales or at least at very large scales, but not within the Solar system.

Two appealing features of our approach stand out. On the theoretical side, CD theory is a paradigm based on a physical interpretation, at least by analogy, of the behavior and properties of space-time and not simply a mathematically consistent frame where facts fit without a corresponding intuitive picture of the situation. On the observational side, the viability of CD is as good as the one of $\Lambda$CDM. The positive tests described in this paper steer us to continue to work out all consequences implied in the Cosmic Defect theory, in quest not only of an alternative picture, but also of some peculiar property allowing really to discriminate one paradigm from the others.

\section*{Acknowledgments}
We would like to thank Alexander Silbergleit and Robert Wagoner of the GP-B theory group at Stanford for stimulating this work.


\end{document}